\documentclass[a4paper,12pt]{article}
\newtheorem{definition}{{\bf Definition}}[section]
\newtheorem{theorem}[definition]{{\bf Theorem}}
\newtheorem{lemma}[definition]{{\bf Lemma}}
\newtheorem{proposition}[definition]{{\bf Proposition}}
\newtheorem{corollary}[definition]{{\bf Corollary}}

\def\<{\langle}
\def\>{\rangle}
\def\im{{\rm i}}

\def\H{{\cal H}}

\def\F{{\cal F}}
\def\card{{\rm card}}

\def\proof{\noindent{Proof. }}
\def\endproof{\hfill $\square$ \vspace{10pt}}
\usepackage{amscd,amsmath,amssymb,graphicx,cite}

\begin{document}

\title{Unitary equivalent classes of one-dimensional quantum walks}
\author{Hiromichi Ohno  \medskip \\
Department of Mathematics, Faculty of Engineering, Shinshu University,\\
4-17-1 Wakasato, Nagano 380-8553, Japan}
\date{}

\maketitle

\begin{abstract}
This study investigates unitary equivalent classes of one-dimensional quantum walks.
We prove that one-dimensional quantum walks are unitary equivalent to
quantum walks of Ambainis type and that 
translation-invariant one-dimensional quantum walks are Szegedy walks.
We also present a necessary and sufficient condition for a one-dimensional quantum walk to
be a Szegedy walk.
\end{abstract}


\section{Introduction}

This study investigates unitary equivalent classes of one-dimensional quantum walks.
A quantum walk is defined by a pair $(U, \{\H_v\}_{v \in V})$, where
$V$ is a countable set, $\{\H_v\}_{v \in V}$ is a family of separable Hilbert spaces,
and $U$ is a unitary operator on $\H = \bigoplus_{v \in V} \H_v$ \cite{SS2}.
For a given quantum walk $(U, \{\H_v\}_{v \in V})$, we can define a graph
$G = (V, D)$ \cite{SS2, HKSS1, HKSS2}.
In this paper, we consider primarily one-dimensional quantum walks,
which have been the subject of many studies
\cite{ADZ, A, ABNVW,EKO, GKD, G, K1,K2, SK, SS, SS2, V, KRBD, K3, KLS}.

It is important to clarify when we think of two quantum walks as being the same.
We consider unitary equivalence of quantum walks in the sense of \cite{SS2}.
If two quantum walks are unitary equivalent, 
then their graphs and dimensions of their Hilbert spaces are the same.
Moreover, the probability distributions of the quantum walks are also the same.
Consequently, we can think of unitary equivalent quantum walks as being the same.

Unitary equivalent classes of simple quantum walks 
have been shown to be 
parameterized by a single-parameter \cite{GKD}.
We extend this result and show that every translation-invariant one-dimensional quantum 
walk is unitary equivalent to a simple quantum walk.
Furthermore, we prove that every one-dimensional quantum walk is 
unitary equivalent to one of Ambainis type.

The Szegedy walk, whose original form was introduced in \cite{Sz}, is 
one of the well-investigated quantum walks (see also \cite{SS,SS2, HKSS3}).
We prove that every translation-invariant one-dimensional quantum walk is a Szegedy walk
and present a necessary and sufficient condition for a one-dimensional quantum walk to
be a Szegedy walk.

The remainder of this paper is organized as follows.
We introduce some notations for quantum walks in Section \ref{sec2}.
In Section \ref{sec3}, we describe the unitary equivalence of quantum walks.
In Section \ref{sec4}, we reveal the form of standard quantum walks.
In Section \ref{sec5}, 
we prove that every one-dimensional quantum walk is 
unitary equivalent to one of Ambainis type.
In Section \ref{sec6}, we clarify when a one-dimensional quantum walk becomes 
a Szegedy walk and show that every translation-invariant one-dimensional quantum walk is
a Szegedy walk.


\section{Preliminaries}\label{sec2}

Let us recall the definition of quantum walks in the sense of \cite{SS,SS2}.

\begin{definition}
Let $V$ be a countable set, $\{\H_v \}_{v \in V}$ a family of separable Hilbert spaces,
and $U$ a unitary on $\H = \bigoplus_{v\in V} \H_v$.
A quantum walk is a pair $(U, \{ \H_v \}_{v \in V})$, and we write
$(U, \{ \H_v \}_{v \in V}) \in \F_{QW}$.
\end{definition}

A (pure) quantum state is represented by a unit vector in a Hilbert space.
For $\lambda \in {\mathbb R}$,
quantum states $\xi$ and $e^{\im \lambda} \xi$ in $\H$ are identified. 
Hence, quantum walks $(U, \{ \H_v \}_{v \in V})$ and $(e^{\im \lambda} U, \{ \H_v \}_{v \in V})$
are also identified.

Let $(U, \{ \H_v \}_{v \in V})$ be a quantum walk.
$P_v \in B(\H)$ is a projection onto $\H_v$, and
$U_{uv} \in B(\H)$ is an operator defined by $U_{uv} = P_u UP_v$
for all $u, v \in V$.
An operator $U_{uv}$ is also considered as an operator in $B(\H_u, \H_v)$, and
we use the same notation if there is no confusion.

Given a quantum walk $(U, \{ \H_v \}_{v \in V}) \in \F_{QW}$,
we can construct a directed graph $G = (V, D)$.
For vertices $u,v \in V$,
the number of directed edges from $v$ to $u$ is denoted by ${\rm card}(u,v)$; i.e.,
\[
{\rm card} (u,v) = {\rm card} \{ e \in D :  t(e)=u, o(e) = v \},
\]
where $o(e)$ and $t(e)$ are the origin and the terminus of the directed edge $e$, respectively,
and card 
indicates the cardinal number of a set.

\begin{definition}
For a quantum walk $(U, \{ \H_v \}_{v \in V}) \in \F_{QW}$,
define the number of directed edges from $v$ to $u$ by
\[
{\rm card} (u,v) = {\rm rank} U_{uv}.
\]
Then a graph $(V,D)$ is called a graph of the quantum walk $(U, \{ \H_v \}_{v \in V})$.
\end{definition}

Next, we define a translation-invariant quantum walk.
Translation-invariant one-dimensional quantum walks are well known.
Here, we extend the notion of translation-invariant quantum walk to arbitrary graphs.

\begin{definition}
A bijection $\gamma$ on $V$ is called an automorphism on a graph $(V, D)$ if 
\[
{\rm card}(u,v) = {\rm card}(\gamma(u), \gamma(v))
\]
for all $u, v \in V$. A quantum walk $(U, \{ \H_v \}_{v \in V})$ is called 
translation invariant for $\gamma$ if
\[
\H_v = \H_{\gamma(v)} \quad \text{and} \quad U_{uv} = U_{\gamma(u) \gamma(v)}
\]
for all $u, v \in V$.
\end{definition}

Now, we introduce three classes of quantum walks.

\begin{definition}
A quantum walk $(U, \{ \H_v \}_{v \in V}) \in \F_{QW}$ is called standard
if the graph is locally finite and symmetric, and satisfies
\[
{\rm card}\{e \in D : o(e) = v \} = \dim \H_v
\]
for all $v \in V$.
\end{definition} 

Note that a symmetric graph satisfies
\[
{\rm card}\{e \in D : o(e) = v \} = {\rm card}\{e \in D : t(e) = v \}.
\]

\begin{definition}
A quantum walk is called one-dimensional if 
$\dim \H_n = 2$, and
the graph of the quantum walk satisfies
$V= {\mathbb Z}$ and
\[
D = \{ (n, n+1), (n+1,n) : n \in {\mathbb Z} \}
\]
with ${\rm card} (n,n+1) ={\rm card} (n+1,n) =1$ for all $n \in {\mathbb Z}$.
\end{definition}

We can canonically define an automorphism $\gamma$ on the graph of 
a one-dimensional quantum walk,
i.e., $\gamma(n) = n+1 $ for $n \in {\mathbb Z}$.

\begin{definition}{\rm \cite{Sz, SS, SS2}}
A standard quantum walk $(U, \{ \H_v \}_{v \in V}) $ is called a Szegedy walk
if there exist a self-adjoint unitary operator $S$ on $\H$, a real number $\lambda \in {\mathbb R}$,
and unit vectors $\phi_v \in \H_v$ such that
$e^{\im \lambda}SU$ has the form
\[
C =\bigoplus_{v \in V}  C_v
\]
where $C_v = 2 | \phi_v \>\< \phi_v | - I_{\H_v}$ on $\H_v$.
Here, the unitary operators $S$ and $C$ are called shift and coin operators, respectively.
\end{definition}

In the case of one-dimensional quantum walks, the operator $C_v$ is a
traceless self-adjoint unitary operator.

Finally, we recall the probability distribution of a quantum walk.

\begin{definition}
Let $(U, \{ \H_v \}_{v \in V}) \in \F_{QW}$, and
let $\Psi_0$ be an initial state in $\H$.
The probability $\mu_t^{\Psi_0} ( v)$ of finding 
the quantum walker at time $t \in {\mathbb Z}_+$ 
and at vertex $v$ is defined by
\[
\mu_t^{\Psi_0} (v) = \| P_v U^t \Psi_0 \|^2.
\]
\end{definition}


\section{Unitary equivalence of quantum walks}\label{sec3}

In this section, we consider the unitary equivalence of quantum walks.

\begin{definition}
$(U_1 , \{ \H_{v_1}^{(1)} \}_{v_1 \in V_1}) \in \F_{QW}$ and 
$(U_2 , \{ \H_{v_2}^{(2)} \}_{v_2 \in V_2}) \in \F_{QW}$ are unitary equivalent,
written $(U_1 , \{ \H_{v_1}^{(1)} \}_{v_1 \in V_1}) \simeq (U_2 , \{ \H_{v_2}^{(2)} \}_{v_2 \in V_2})$,
if there exist a unitary $W$ from $\bigoplus_{v_1\in V_1} \H_{v_1}$ to 
$\bigoplus_{v_2\in V_2} \H_{v_2}$ and a bijection $\phi$ from $V_1$ to $V_2$ such that
\[
W \H_{v_1} = \H_{\phi(v_1)}  \quad \text{and} \quad W U_1 W^* = U_2.
\]
\end{definition}

We would like to regard unitary equivalent quantum walks as being the same.
The next proposition says that unitary equivalent quantum walks have the same graphs.

\begin{proposition}\label{prop3.2}
Let $(U_1 , \{ \H_{v_1}^{(1)} \}_{v_1 \in V_1})$ and 
$(U_2 , \{ \H_{v_2}^{(2)} \}_{v_2 \in V_2})$ be quantum walks, and let 
$G_1 = (V_1, D_1)$ and $G_2 = (V_2, D_2)$ be their graphs.
If quantum walks $(U_1 , \{ \H_{v_1}^{(1)} \}_{v_1 \in V_1})$ and 
$(U_2 , \{ \H_{v_2}^{(2)} \}_{v_2 \in V_2})$ are unitary equivalent by a unitary $W$,
the graphs $G_1 = (V_1, D_1)$ 
and $G_2 = (V_2, D_2)$ are isomorphic; that is,
there exists a bijection $\phi$ from $V_1$ to $V_2$ such that
\[
{\rm card} (u,v) = {\rm card} (\phi(u), \phi(v)).
\]
\end{proposition}

\proof
From the definition of unitary equivalence, there exists a bijection $\phi$ from $V_1$ to $V_2$.
A unitary $W$ maps $\H_{v_1}$ to $\H_{\phi(v_1)}$,
with the result that $W P_{v_1} W^* = P_{\phi(v_1)}$.
Since $\card (u,v) = {\rm rank} P_u  U_1 P_v$ for $u,v \in V_1$,
\[
\card (u,v) =  {\rm rank} P_u U_1 P_v =   {\rm rank} W P_u U_1 P_v W^*
=  {\rm rank} P_{\phi(u)} U_2 P_{\phi(v)} =\card (\phi(u),\phi(v)). 
\]
Hence, we obtain the proposition. 
\endproof

When $(U_1 , \{ \H_{v_1}^{(1)} \}_{v_1 \in V_1}) \in \F_{QW}$ and 
$(U_2 , \{ \H_{v_2}^{(2)} \}_{v_2 \in V_2}) \in \F_{QW}$ are unitary equivalent,
we can identify $V_2$ with $V_1$ using the bijection $\phi$, and write $V = V_1$.
Similarly, $D_1$ and $D_2$, and $\H^{(1)}_{v}$ and $\H^{(2)}_{\phi(v)}$ can be
identified, and we write $D= D_1$ and $\H_v = \H_{v}^{(1)}$. 
Here, the unitary $W$ can be decomposed as
\[
W = \bigoplus_{v\in V} W_v,
\]
where $W_v = P_v W P_v$.

The following corollary is an immediate consequence of Proposition \ref{prop3.2}.

\begin{corollary}
Let quantum walks $(U_1, \{ \H_v \}_{v\in V})$ and  $(U_2, \{ \H_v \}_{v\in V})$ be
unitary equivalent.
If $(U_1, \{ \H_v \}_{v\in V})$ is standard or one-dimensional, 
then so is $(U_2, \{ \H_v \}_{v\in V})$.
\end{corollary}

Unitary equivalence also preserves the properties of a Szegedy walk.

\begin{proposition}\label{prop3.4}
Let quantum walks $(U_1, \{ \H_v \}_{v\in V})$ and  $(U_2, \{ \H_v \}_{v\in V})$ be
unitary equivalent by a unitary $W$.
If $(U_1, \{ \H_v \}_{v\in V})$ is a Szegedy walk, then so is $(U_2, \{ \H_v \}_{v\in V})$.
\end{proposition}

\proof
By the assumption, there exist a self-adjoint unitary $S$ on $\H$, 
a real number $\lambda \in {\mathbb R}$,  and
unit vectors $\phi_v \in \H_v$ such that $e^{\im \lambda} SU_1$ has the form
\[
C = \bigoplus_{v \in V} C_v,
\]
where $C_v = 2 |\phi_v\>\< \phi_v|-I_{\H_v}$.
Then, $WSW^*$ is also a self-adjoint unitary on  $\H$.
Moreover,
\[
e^{\im \lambda} WSW^*U_2 = e^{\im \lambda} WSU_1W^* = W \bigoplus_{v\in V} C_v W^* 
= \bigoplus_{v\in V} 2|W\phi_v \>\< W\phi_v| -I_{H_v},
\]
from which it follows that $(U_2, \{ \H_v \}_{v\in V})$ is a Szegedy walk.
\endproof

In general, a quantum walk that is unitary equivalent to a 
translation-invariant quantum walk is not translation invariant.
However, if we add a condition, then translation-invariance is also preserved.

\begin{proposition}\label{prop3.5}
Let  $(U, \{ \H_n \}_{n\in {\mathbb Z}})$ be a translation-invariant one-dimensional
quantum walk such that  $\H_n =\H_{n+1}$ for all $n \in {\mathbb Z}$,
 and let $W$ be a unitary on $\H$ that has the form
\[
W = \bigoplus_{n\in{\mathbb Z}} W_n,
\]
where $W_n$ is a unitary on $\H_n$, and $W_n = W_{n+1}$ for all $n \in {\mathbb Z}$.
The quantum walk $(WUW^*, \{ \H_n \}_{n\in {\mathbb Z}})$ is a translation-invariant
 one-dimensional
quantum walk.
\end{proposition}

\proof
It is sufficient to prove that $WUW^*$ is translation invariant; i.e.,
$P_n WUW^* P_m = P_{\gamma(n)} WUW^* P_{\gamma(m)}$.
Since $U$ is translation invariant,
\begin{eqnarray*}
P_n WUW^* P_m &=& W_nP_nUP_mW_m^* =  W_n U_{nm}W_m^*
= W_{\gamma(n)} U_{\gamma(n) \gamma(m)}W_{\gamma(m)}^* \\
&=& W_{\gamma(n)} P_{\gamma(n)} U P_{\gamma(m)} W_{\gamma(m)}^* 
=P_{\gamma(n)} WUW^* P_{\gamma(m)}.
\end{eqnarray*}
Hence, $WUW^*$ is translation invariant.
\endproof

Finally, we consider the probability distribution of a quantum walk.
This does not change under unitary equivalence.

\begin{proposition}
Let quantum walks $(U_1, \{ \H_v \}_{v\in V})$ and  $(U_2, \{ \H_v \}_{v\in V})$ be
unitary equivalent by a unitary $W$,
let $\Phi_0$ and $W\Phi_0$ in $\H$ be an initial state of $(U_1, \{ \H_v \}_{v\in V}) $
and $(U_2, \{ \H_v \}_{v\in V})$, respectively, 
and let $\mu_t^{(1), \Phi_0}$ and $\mu_t^{(2), W\Phi_0}$ be 
the probability distributions of the quantum walks $(U_1, \{ \H_v \}_{v\in V}) $ and
$(U_2, \{ \H_v \}_{v\in V}) $, respectively.
Then, 
\[
\mu_t^{(1), \Phi_0}(v) =  \mu_t^{(2), W\Phi_0}(v)
\]
for all $t \in {\mathbb Z}+$ and $v \in V$.
\end{proposition}

\proof
By the definition,
\[
 \mu_t^{(2), W\Phi_0}(v) = \| P_v (WU_1W^*)^t W \Phi_0 \|^2 = \|W P_v U_1^t \Phi_0 \|^2
= \| P_v U_1^t \Phi_0 \|^2 =\mu_t^{(1), \Phi_0}(v).
\]
Therefore, we obtain the proposition.
\endproof

One of the primary topics of study in connection with quantum walks 
is the probability distributions of quantum walks,
by virtue of which we can think of unitary equivalent quantum walks as being the same.
When we consider other properties of quantum walks, 
additional properties, such as Proposition \ref{prop3.5}, must be considered.


\section{Standard quantum walk}\label{sec4}

This study investigates one-dimensional quantum walks and Szegedy walks.
Since both kinds of quantum walk are standard,
we clarify the form of a standard quantum walk.

\begin{theorem}\label{thm4.1}
Let $(U, \{ \H_v \}_{v \in V}) \in \F_{QW}$ be a standard quantum walk.
There exist orthonormal bases $\{\xi_{e}\}_{e \in D}$ 
and $\{\zeta_{e}\}_{e\in D}$ of $\H$ with
$\xi_{e} \in \H_{t(e)}$ and $\zeta_{e} \in \H_{o(e)}$, such that
\[
U = \sum_{e \in D} |\xi_e\> \< \zeta_e|.
\]
Moreover, $U_{uv}$ has the form
\[
U_{uv} = \sum_{e : t(e) =u, o(e)=v} |\xi_{e} \>\< \zeta_{e} |
\]
for any $u, v \in V$.
\end{theorem}

\proof
Since ${\rm rank}U_{uv} = {\rm card} \{ e \in V : t(e) = u, o(e) = v\}$, 
we can set $\{\xi_e : t(e) = u, o(e) =v, e\in D \} \subset \H_u = \H_{t(e)}$ 
as an orthonormal basis of ${\rm ran} U_{uv}$ for all $u,v \in V$.
Then, $\{\xi_e : o(e) =v, e\in D \}$ is an orthonormal system of $\H$.
An operator $UP_v$ is a partial isometry with an initial projection $P_v$.
The range of this operator is contained in 
$\bigoplus_{u\in V} {\rm ran} U_{uv} = {\rm span}\{ \xi_{e} : o(e) =v , e \in D \}$, that is,
\begin{equation}\label{4.1.1}
{\rm ran} U P_v \subset  {\rm span}\{ \xi_{e} : o(e) =v , e \in D \}.
\end{equation}
From the definition of a standard quantum walk,
$ \dim \H_v = {\rm card} \{ e \in D : o(e) = v\}$.
Since the rank of the range projection of $U P_v$ is equal to the rank of the initial projection $P_v$,
${\rm rank}  U P_v = {\rm card} \{ e \in D : o(e) = v\}$.
Considering the dimensions of the subspaces in \eqref{4.1.1}, 
\[
{\rm ran}UP_v = {\rm span}\{ \xi_{e} : o(e) =v , e \in D \}.
\]
Moreover, the range projection $UP_v U^*$ leaves 
${\rm span}\{ \xi_{e} : o(e) =v , e \in D \}$ unchanged.
Therefore, $U P_v U^* \xi_e = \xi_e$ for all $e \in D$ with $o(e) =v$, and
this implies that $P_v U^*  \xi_e  = U^* \xi_e$, with the result that $U^* \xi_e \in \H_v =\H_{o(e)}$.

Let $\zeta_{e} = U^* \xi_{e}$. Since $\{ \xi_{e} : o(e) = v , e\in D \}$ is an orthonormal system,
$\{\zeta_{e} : o(e) =v , e\in D \}$ is an orthonormal basis of $\H_v$.
Hence, $\{ \zeta_{e} : e \in D\}$ is an orthonormal basis of $\H$.
Since $U$ is unitary and $U\zeta_{e} = \xi_{e}$, 
$\{\xi_{e} : e \in D \}$ is also an orthonomal basis of $\H$, and
\[
U = \sum_{e\in D} |\xi_{e} \>\< \zeta_{e}|.
\]
By the definition of $U_{uv}$, the equation
\[
U_{uv} = \sum_{e : t(e) =u, o(e)=v} |\xi_{e} \>\< \zeta_{e} |
\]
also holds.
\endproof

\begin{corollary}
For a standard quantum walk $(U, \{ \H_v \}_{v \in V}) \in \F_{QW}$,
there exist a self-adjoint unitary operator $S$ on $\H$ and 
a unitary operator $T_v$ on $ \H_v$ $(v \in V)$
such that $U = ST$, where $T = \bigoplus_{v \in V} T_v$.
\end{corollary}

\proof
Since the graph of a standard quantum walk is symmetric,
there exists a bijection on $D$, denoted by $e \mapsto \bar{e}$, for which
$t(e) = o(\bar{e})$, $o(e) = t(\bar{e})$, and $\bar{\bar{e}} =e$.

By Theorem \ref{thm4.1}, $U$ can be written as
\[
U = \sum_{e \in D} |\xi_{e}\> \< \zeta_{e}|.
\]
Define $S$ by $S\xi_{e} = \xi_{\bar{e}}$. $S$ is a self-adjoint unitary; hence, $S^2=I$.
Then, 
\[
SU = \sum_{e\in D} |\xi_{\bar{e}} \>\< \zeta_{e}| =
\bigoplus_{v \in V} \sum_{o(e) =v }  |\xi_{\bar{e}} \>\< \zeta_{e}|.
\]
The operator 
\[
T_v=\sum_{o(e) =v }  |\xi_{\bar{e}} \>\< \zeta_{e}|
\]
satisfies the assertion.
\endproof

Now, to clarify the explicit form of a shift operator $S$ of 
a Szegedy walk,
we present the next lemma.

\begin{lemma}\label{lem4.3}
Let $(U, \{\H_v\}_{v\in V})$ be a Szegedy walk with a shift operator $S$ and a coin operator 
$C$,
such that $U = e^{\im \lambda} SC$ for some $\lambda \in {\mathbb R}$.
Then,
\[
S({\rm ran} U_{uv}) ={\rm ran} U_{vu}
\]
for any $u, v \in V$.
\end{lemma}
\proof
By Theorem \ref{thm4.1}, we can assume that
there exist orthonormal bases $\{\xi_{e}\}_{e \in D}$ 
and $\{\zeta_{e}\}_{e\in D}$ of $\H$ with
$\xi_{e} \in \H_{t(e)}$ and $\zeta_{e} \in \H_{o(e)}$, such that
\[
U = \sum_{e \in D} |\xi_e\> \< \zeta_e|.
\]
Then, ${\rm ran} U_{uv} = {\rm span} \{ \xi_e: t(e) = u, o(e) =v, e\in D \}$.
Since the coin operator $C$ is written as a direct sum of $C_v$, 
\[
S \xi_e = SU \zeta_e = e^{\im \lambda} C \zeta_e \in \H_v = \H_{o(e)}
\]
for all $e \in D$ with $t(e)=u$ and $o(e) = v$.
This implies that $S({\rm ran} U_{uv}) \subset \H_v$.
Furthermore, by the form of $U$, $\H_v$ is decomposed as
\[
\H_v = \bigoplus_{w \in V} {\rm ran} U_{vw}.
\]
Here,
\[
{\rm ran} U_{uv} = S^2 ( {\rm ran} U_{uv}) \subset S \H_v  = \bigoplus_{w \in V} S( {\rm ran} U_{vw}).
\]
Since $S( {\rm ran} U_{vw}) \subset \H_w$ and ${\rm ran} U_{uv} \subset \H_u$,  
${\rm ran} U_{uv} \subset S({\rm ran} U_{vu})$. 
Considering the inversion formula,
\[
S({\rm ran} U_{uv}) = {\rm ran} U_{vu}
\]
for all $u,v \in V$.
\endproof

Using this lemma, we have the next theorem.

\begin{theorem}
Let $(U, \{\H_v\}_{v\in V})$ be a Szegedy walk with a shift operator $S$ and a coin operator 
$C$,
such that $U = e^{\im \lambda} SC$ for some $\lambda \in {\mathbb R}$.
There exist orthonormal bases $\{\xi_e \}_{e \in D}$ 
and $\{\zeta_e \}_{e \in D}$ of $\H$ with $\xi_e \in \H_{t(e)}$ and $\zeta_e \in \H_{o(e)}$
such that 
\[
U = \sum_{e\in D} |\xi_e\>\< \zeta_e| \quad {\rm and } \quad
S = \sum_{e \in D} |\xi_e\>\<\xi_{\bar{e}}|.
\]
\end{theorem}

\proof
By Theorem \ref{thm4.1}, we can assume that
there exist orthonormal bases $\{\xi_{e}\}_{e \in D}$ 
and $\{\zeta_{e}\}_{e\in D}$ of $\H$ with
$\xi_{e} \in \H_{t(e)}$ and $\zeta_{e} \in \H_{o(e)}$ such that
\[
U = \sum_{e \in D} |\xi_e\> \< \zeta_e|.
\]
By Lemma \ref{lem4.3}, $S({\rm ran} U_{uv}) = {\rm ran} U_{vu}$.
Moreover, in the proof of Theorem \ref{thm4.1}, the choice of an orthonormal basis of 
${\rm ran} U_{uv}$ is arbitrary.
Therefore, for an orthonormal basis $\{\xi_e : t(e)=u, o(e) =v , e\in D \}$ of ${\rm ran}U_{uv}$,
we can redefine $\xi_{\bar{e}} = S \xi_e$. Then, $\{ \xi_{\bar{e}} : t(e)=u, o(e) =v , e\in D \}$
is an orthonormal basis of ${\rm ran}U_{vu}$.
Consequently, we can obtain orthonormal bases 
$\{\xi_e \}_{e \in D}$ and $\{\zeta_e \}_{e \in D}$ of $\H$ 
with $\xi_e \in \H_{t(e)}$ and $\zeta_e \in \H_{o(e)}$
such that
\[
U = \sum_{e\in D} |\xi_e\>\< \zeta_e| \quad {\rm and } \quad
S = \sum_{e \in D} |\xi_e\>\<\xi_{\bar{e}}|.
\]


\section{One-dimensional quantum walks}\label{sec5}

In this section, we consider a one-dimensional quantum walk
$(U, \{ \H_n \}_{n\in {\mathbb Z}})$.
Without loss of generality, we can assume that $\H_n = {\mathbb C}^2$ for all $n \in {\mathbb Z}$.
Here, $D = \{ (n, n+1) , (n+1,n) : n \in {\mathbb Z}\}$.
By Theorem \ref{thm4.1}, there exist orthonormal bases 
$\{ \xi_{n,n+1} ,\xi_{n+1,n} \}_{n \in {\mathbb Z}}$ and
$\{ \zeta_{n,n+1}, \zeta_{n+1,n} \}_{n \in{\mathbb Z}}$ of $\H$ with 
$\xi_{n,n+1},\zeta_{n+1,n} \in \H_n$ and $\xi_{n+1,n} , \zeta_{n,n+1} \in \H_{n+1}$
such that
\[
U= \sum_{n \in {\mathbb Z}} |\xi_{n,n+1}\> \< \zeta_{n,n+1}| + |\xi_{n+1,n}\> \<\zeta_{n+1,n}| 
=\sum_{n \in {\mathbb Z}} |\xi_{n-1,n} \>\< \zeta_{n-1,n}| + |\xi_{n+1,n} \>\<\zeta_{n+1,n}|.
\]

There is a substantial literature on one-dimensional quantum walks,
which fall into four principal types.
The first type is represented as follows. 
Let $\{e_1^n, e_2^n\}$ be a canonical orthonormal basis of $\H_n = {\mathbb C}^2$.
We consider $e_i^n$ as $|n\>|i\>$.
We take $\xi_{n-1,n} = e_1^{n-1}$, $\xi_{n+1,n} = e_2^{n+1}$, and 
$\zeta_{n-1, n} = \bar{a}_n e_1^n + \bar{b}_n e_2^n$,
$\zeta_{n+1, n} = \bar{c}_n e_1^n + \bar{d}_n e_2^n$.
Then,
\begin{eqnarray*}
U_{n-1,n} &=& |e_1^{n-1}\>\<\bar{a}_n e_1^n + \bar{b}_n e_2^n |
= |n-1\>\<n| \otimes \left[ \begin{matrix} a_n & b_n \\ 0 & 0 \end{matrix} \right] \\
U_{n+1,n} &=& |e_2^{n+1}\>\<\bar{c}_n e_1^n + \bar{d}_n e_2^n |
= |n+1\>\<n| \otimes \left[ \begin{matrix} 0 & 0 \\ c_n & d_n \end{matrix} \right].
\end{eqnarray*}
A quantum walk of this type is said to be of Ambainis type \cite{A, ABNVW}.
A set of all quantum walks of this type is denoted by ${\cal C}_1$.
Note that 
\[
\left[ \begin{matrix} a_n & b_n \\ c_n & d_n \end{matrix} \right] 
\]
is unitary.

The second type is represented by taking
$\xi_{n-1, n} = {a}_{n} e_1^{n-1} + c_n e_2^{n-1}$, 
$\xi_{n+1, n} = {b}_n e_1^{n+1} + {d}_n e_2^{n+1}$, and 
$\zeta_{n-1,n} = e_1^{n}$, $\zeta_{n+1,n} = e_2^{n}$, such that 
\begin{eqnarray*}
U_{n-1,n} &=& | {a}_{n} e_1^{n-1} + c_n e_2^{n-1}\>\<e_1^{n} |
= |n-1\>\<n| \otimes \left[ \begin{matrix} a_n & 0 \\ c_n & 0 \end{matrix} \right] \\
U_{n+1,n} &=& | {b}_n e_1^{n+1} + {d}_n e_2^{n+1}\>\< e_2^{n} |
= |n+1\>\<n| \otimes \left[ \begin{matrix} 0 & b_n \\ 0 & d_n \end{matrix} \right].
\end{eqnarray*}
A quantum walk of this type is said to be of Gudder type \cite{G}.
A set of all quantum walks of this type is denoted by ${\cal C}_2$.

Similarly, the third type is represented by taking
$\xi_{n-1,n} = e_2^{n-1}$, $\xi_{n+1,n} = e_1^{n+1}$, and 
$\zeta_{n-1, n} = \bar{c}_n e_1^n + \bar{d}_n e_2^n$,
$\zeta_{n+1, n} = \bar{a}_n e_1^n + \bar{b}_n e_2^n$, such that 
\begin{eqnarray*}
U_{n-1,n} &=& |e_2^{n-1}\>\<\bar{c}_n e_1^n + \bar{d}_n e_2^n |
= |n-1\>\<n| \otimes \left[ \begin{matrix} 0 & 0 \\ c_n & d_n \end{matrix} \right] \\
U_{n+1,n} &=& |e_1^{n+1}\>\<\bar{a}_n e_1^n + \bar{b}_n e_2^n |
= |n+1\>\<n| \otimes \left[ \begin{matrix} a_n & b_n \\ 0 & 0\end{matrix} \right].
\end{eqnarray*}
A set of all quantum walks of this type is denoted by ${\cal C}_3$.
The fourth type is
represented by taking
$\xi_{n-1, n} = {b}_{n} e_1^{n-1} + d_n e_2^{n-1}$, 
$\xi_{n+1, n} = {a}_n e_1^{n+1} + {c}_n e_2^{n+1}$, and 
$\zeta_{n-1,n} = e_2^{n}$, $\zeta_{n+1,n} = e_1^{n}$, such that
\begin{eqnarray*}
U_{n-1,n} &=& | {b}_{n} e_1^{n-1} + d_n e_2^{n-1}\>\<e_2^{n} |
= |n-1\>\<n| \otimes \left[ \begin{matrix} 0  & b_n \\ 0 & d_n \end{matrix} \right] \\
U_{n+1,n} &=& | {a}_n e_1^{n+1} + {c}_n e_2^{n+1}\>\< e_1^{n} |
= |n+1\>\<n| \otimes \left[ \begin{matrix} a_n &0 \\ c_n & 0  \end{matrix} \right].
\end{eqnarray*}
A set of all quantum walks of this type is denoted by ${\cal C}_4$.

Summarizing, we have four types of one-dimensional quantum walks:
\begin{eqnarray*}
\text{(1)} && U_{n-1, n} =  |n-1\>\<n| \otimes
\left[ \begin{matrix} a_n & b_n \\ 0 & 0 \end{matrix} \right], \quad
U_{n+1, n} =  |n+1\>\<n| \otimes \left[ \begin{matrix} 0 & 0 \\ c_n & d_n \end{matrix} \right], \\
\text{(2)} && U_{n-1, n} =  |n-1\>\<n| \otimes
\left[ \begin{matrix} a_n & 0 \\ c_n & 0 \end{matrix} \right], \quad
U_{n+1, n} =  |n+1\>\<n| \otimes \left[ \begin{matrix} 0 & b_n \\ 0 & d_n \end{matrix} \right], \\
\text{(3)} && U_{n-1, n} =  |n-1\>\<n| \otimes
\left[ \begin{matrix} 0 & 0 \\ c_n & d_n \end{matrix} \right], \quad
U_{n+1, n} =  |n+1\>\<n| \otimes \left[ \begin{matrix} a_n & b_n \\ 0 & 0 \end{matrix} \right], \\
\text{(4)} && U_{n-1, n} =  |n-1\>\<n| \otimes
\left[ \begin{matrix} 0 & b_n \\ 0 & d_n \end{matrix} \right], \quad
U_{n+1, n} = |n+1\>\<n| \otimes  \left[ \begin{matrix} a_n & 0 \\ c_n & 0 \end{matrix} \right] .
\end{eqnarray*}
These four types of one-dimensional quantum walks are also represented as
follows:
\begin{eqnarray*}
\text{(1)} && 
U = \sum_{n \in {\mathbb Z}} |e^{n-1}_1 \> \< \zeta_{n-1,n}| + |e^{n+1}_2 \> \< \zeta_{n+1,n} |,\\
\text{(2)} && 
U = \sum_{n \in {\mathbb Z}} |\xi_{n-1,n} \> \< e^{n}_1| + |\xi_{n+1,n} \> \< e^{n}_2 |, \\
\text{(3)} && 
U = \sum_{n \in {\mathbb Z}} |e^{n-1}_2 \> \< \zeta_{n-1,n}| + |e^{n+1}_1 \> \< \zeta_{n+1,n} |,\\
\text{(4)} && 
U = \sum_{n \in {\mathbb Z}} |\xi_{n-1,n} \> \< e^{n}_2| + |\xi_{n+1,n} \> \< e^{n}_1 |.
\end{eqnarray*}

\begin{theorem}\label{thm5.2}
Let $(U, \{ \H_n \}_{n \in {\mathbb Z}})$ be  a one-dimensional quantum walk.
For each $k =1,2,3,4$,
there exists a one-dimensional quantum walk in ${\cal C}_k$
 that is unitary equivalent to $(U, \{ \H_n \}_{n \in {\mathbb Z}})$.
\end{theorem}

\proof
By Theorem \ref{thm4.1}, $U$ can be written as
\[
U = \sum_{n \in {\mathbb Z}} |\xi_{n-1,n} \>\< \zeta_{n-1,n}| + |\xi_{n+1,n} \>\<\zeta_{n+1,n}|,
\]
where
$\{ \xi_{n,n+1} ,\xi_{n+1,n} \}_{n \in {\mathbb Z}}$ and
$\{ \zeta_{n,n+1}, \zeta_{n+1,n} \}_{n \in{\mathbb Z}}$ are orthonormal bases of $\H$ with 
$\xi_{n,n+1},\zeta_{n+1,n} \in \H_n$ and $\xi_{n+1,n} , \zeta_{n,n+1} \in \H_{n+1}$.

First, we prove that 
$(U, \{ \H_n \}_{n \in {\mathbb Z}})$ is
 unitary equivalent to a quantum walk in  ${\cal C}_1$.
Let 
\[
W_n =  |e_1^n \> \< \xi_{n,n+1}| +  | e_2^n\>\< \xi_{n,n-1} |
\]
for all $n \in {\mathbb Z}$.
It is easily seen that $W_n$ is a unitary on $\H_n$, with the result that
$W = \bigoplus_{n\in {\mathbb Z}} W_n$ is a unitary on $\H$ that satisfies $W\H_n = \H_n$.
Moreover, from a direct calculation,
\begin{eqnarray*}
WUW^* &=& W \left( \sum_{n \in {\mathbb Z}} |\xi_{n-1,n} \>\< \zeta_{n-1,n}| 
+ |\xi_{n+1,n} \>\<\zeta_{n+1,n}| \right) W^*  \\
&=&  
\sum_{n \in {\mathbb Z}} |e^{n-1}_1 \> \< W\zeta_{n-1,n}| + |e^{n+1}_2 \> \< W \zeta_{n+1,n} |.
\end{eqnarray*}
Since $W= \bigoplus_{n\in {\mathbb Z}} W_n$  is unitary,
$\{W\zeta_{n-1,n} , W\zeta_{n+1,n}\}$ is an orthonormal basis of $\H_n$.
Hence, $(WUW^*, \{\H_n\}_{n\in{\mathbb Z}})$ is in ${\cal C}_1$, and
we obtain that $(U, \{ \H_n \}_{n \in {\mathbb Z}})$ is
unitary equivalent to a quantum walk in  ${\cal C}_1$.

Second, we prove that $(U, \{ \H_n \}_{n \in {\mathbb Z}})$ is
unitary equivalent to a quantum walk  in  ${\cal C}_2$.
Let 
\[
W_n =   |  e_1^n\>\< \zeta_{n-1,n} | + | e_2^n\> \< \zeta_{n+1,n} |
\]
for all $n \in {\mathbb Z}$.
It is easily seen that $W_n$ is a unitary on $\H_n$, with the result that
$W = \bigoplus_{n\in {\mathbb Z}} W_n$ is a unitary on $\H$.
Moreover, from a direct calculation, we have
\begin{eqnarray*}
WUW^* &=&
W \left(  \sum_{n \in {\mathbb Z}} |\xi_{n-1,n} \>\< \zeta_{n-1,n}| + 
|\xi_{n+1,n} \>\<\zeta_{n+1,n}|  \right)
 W^* \\
&=& 
 \sum_{n \in {\mathbb Z}} | W \xi_{n-1,n} \> \< e^{n}_1| + |W \xi_{n+1,n} \> \< e^{n}_2 |.
\end{eqnarray*}
Since $W= \bigoplus_{n\in {\mathbb Z}} W_n$  is unitary,
$\{W\xi_{n,n-1} , W\xi_{n,n+1}\}$ is an orthonormal basis of $\H_n$.
Hence, $(WUW^*, \{\H_n\}_{n\in{\mathbb Z}})$ is in ${\cal C}_2$, and
we obtain that $(U, \{ \H_n \}_{n \in {\mathbb Z}})$ is
unitary equivalent to a quantum walk  in  ${\cal C}_2$.

The proofs of the remaining parts are similar to these.
\endproof

As a corollary of the theorem,
we have the following.

\begin{corollary}
Let $(U, \{ \H_n \}_{n \in {\mathbb Z}})$ be  a translation-invariant one-dimensional quantum walk.
For each $k =1,2,3,4$,
there exists a translation-invariant one-dimensional quantum walk in ${\cal C}_k$
 that is unitary equivalent to $(U, \{ \H_n \}_{n \in {\mathbb Z}})$.
\end{corollary}

\proof
By Theorem \ref{thm4.1}, $U$ can be written as
\[
U = \sum_{n \in {\mathbb Z}} |\xi_{n-1,n} \>\< \zeta_{n-1,n}| + |\xi_{n+1,n} \>\<\zeta_{n+1,n}|,
\]
where
$\{ \xi_{n,n+1} ,\xi_{n+1,n} \}_{n \in {\mathbb Z}}$ and
$\{ \zeta_{n,n+1}, \zeta_{n+1,n} \}_{n \in{\mathbb Z}}$ are orthonormal bases of $\H$ with 
$\xi_{n,n+1},\zeta_{n+1,n} \in \H_n$ and $\xi_{n+1,n} , \zeta_{n,n+1} \in \H_{n+1}$.
Moreover, by translation invariance, we can assume that
\[
\xi_{n,n+1} = \xi_{n-1,n}, \quad \xi_{n+1,n} = \xi_{n,n-1}, \quad
\zeta_{n,n+1} = \zeta_{n-1,n} \quad {\rm and } \quad \zeta_{n+1,n} = \zeta_{n,n-1}
\]
for all $n \in {\mathbb Z}$.
Therefore, in the proof of Theorem \ref{thm5.2}, $W_n = W_{n+1}$ for all $n \in {\mathbb Z}$.
Then, the assertion of the corollary follows from Proposition \ref{prop3.5}.
\endproof


\section{One-dimensional Szegedy walk}\label{sec6}

In this section, we consider a necessary and sufficient condition for 
a one-dimensional quantum walk to be a Szegedy walk.
Let $(U, \{\H_n \}_{n\in {\mathbb Z}})$ be a one-dimensional quantum walk.
Considering the unitary equivalence, 
we can assume $\H_n = {\mathbb C}^2$ for all $n \in {\mathbb Z}$ without loss of generality.
By theorem \ref{thm5.2}, we can assume that $U$ is represented as
\[
U =\sum_{n \in {\mathbb Z}} |e^{n-1}_1 \> \< \zeta_{n-1,n}| + |e^{n+1}_2 \>\< \zeta_{n+1,n} |,
\]
where $\{\zeta_{n-1,n} , \zeta_{n+1,n}\}$ is an orthonormal basis of $\H_n$.
Here, 
\begin{eqnarray*}
U_{n-1, n} &=&  |e^{n-1}_1 \> \< \zeta_{n-1,n}| =
|n-1\> \<n| \otimes \left[ \begin{matrix} a_n & b_n \\ 0 & 0 \end{matrix} \right], \\
U_{n+1, n} &=& |e^{n+1}_2 \> \< \zeta_{n+1,n} |= |n+1\>\<n| \otimes 
\left[ \begin{matrix} 0 & 0 \\ c_n & d_n \end{matrix} \right],
\end{eqnarray*}
where the matrix
\[
\left[ \begin{matrix} a_n & b_n \\ c_n & d_n \end{matrix} \right]
\]
is unitary for all $n \in {\mathbb Z}$.

If this is a Szegedy walk, there exists a shift operator $S$ such that
$e^{\im \lambda} SU$ is a direct sum of traceless self-adjoint unitary operators 
for some $\lambda \in {\mathbb R}$.
By Lemma \ref{lem4.3}, 
 $S({\rm ran}U_{n,n+1}) = {\rm ran}U_{n+1, n}$. 
Moreover, ${\rm ran}U_{n+1,n} = {\mathbb C} e_2^{n+1}$ and 
 ${\rm ran}U_{n, n+1} = {\mathbb C} e_1^{n}$. 
Therefore, $S e_1^n =e^{\im \theta_n} e_2^{n+1}$ for some $\theta_n \in {\mathbb R}$.
Consequently, $S$ has the form
\begin{equation}\label{eq6.1.1b}
S= \sum_{n\in{\mathbb Z}} e^{\im \theta_n} |e^{n+1}_{2}\> \<e^{n}_{1} | 
+ e^{-\im \theta_n} |e^{n}_{1}\>\<e^{n+1}_{2} |.
\end{equation}
Then, $SU$ is described as
\[
SU = \bigoplus_{n \in {\mathbb Z}} 
\left[ \begin{matrix} e^{-\im \theta_{n}} c_n & e^{-\im \theta_{n}} d_n \\ 
 e^{\im \theta_{n-1}}  a_n &  e^{\im \theta_{n-1}} b_n \end{matrix} \right].
\]
Let $c_n = e^{\im \mu_n} r_n$ and $b_n = e^{\im \nu_n} r_n$ with $r_n \ge 0$ 
and $\mu_n, \nu_n \in  {\mathbb R}$.
Then,
\begin{equation}\label{eq6.1.1}
e^{\im \lambda} SU = 
\bigoplus_{n \in {\mathbb Z}}
\left[ \begin{matrix} e^{\im (-\theta_{n} +\lambda +\mu_n)} r_n
& e^{\im (-\theta_{n}+\lambda)} d_n \\ 
 e^{\im (\theta_{n-1}+ \lambda)}  a_n &  e^{\im (\theta_{n-1} + \lambda +\nu_n)} r_n \end{matrix} \right].
\end{equation}
When $r_n \neq 0$, the $2\times 2$ matrices on the right hand side 
are traceless self-adjoint unitary if and only if
\begin{equation}\label{eq6.1.2}
-\theta_{n} +\lambda +\mu_n =0 \quad ({\rm mod} \  \pi),  \qquad
- \theta_{n} +\mu_n = \theta_{n-1} + \nu_n + \pi \qquad  ({\rm mod} \  2\pi) .
\end{equation}
In the case $r_n = 0$, let $a_n = e^{\im \sigma_n}$ and $d_n = e^{\im \tau_n}$
for some $\sigma_n, \tau_n \in  {\mathbb R}$.
Then the  $2\times 2$ matrices on the right hand side in \eqref{eq6.1.1}
are traceless self-adjoint unitary  if and only if
\begin{equation}\label{eq6.1.3}
\theta_{n-1} + \lambda + \sigma_n = \theta_{n} -\lambda -\tau_n \qquad  ({\rm mod} \  2\pi) .
\end{equation}
Hence, $\theta_n$ and $\lambda$ satisfy conditions \eqref{eq6.1.2} and \eqref{eq6.1.3}.

Conversely, if there exist $\theta_n$ and $\lambda$ satisfying these conditions,
the quantum walk $(U, \{\H_n \}_{n \in {\mathbb Z}})$ is a Szegedy walk.
Indeed, define a shift operator $S$ by \eqref{eq6.1.1b}. 
Then,  it is easily seen that $e^{\im \lambda} SU$ is a direct sum of 
traceless self-adjoint unitary operators.

Therefore, $(U, \{\H_n \}_{n \in {\mathbb Z}})$ is a Szegedy walk if and only if the above 
simultaneous equations for $\lambda$ and $\theta_n$
have a solution.

Now, we have the next theorem.

\begin{theorem}\label{thm6.1}
Let  $(U, \{\H_n \}_{n \in {\mathbb Z}})$ be a one-dimensional quantum walk given by
\begin{eqnarray}
U_{n-1, n} &=&  |e^{n-1}_1 \> \< \zeta_{n-1,n}| = |n-1\>\<n| \otimes
\left[ \begin{matrix} e^{\im \sigma_n} s_n & e^{\im \nu_n} r_n \\ 0 & 0 \end{matrix} \right], 
\nonumber \\
U_{n+1, n} &=&  |e^{n+1}_2 \> \< \zeta_{n+1,n} |= |n+1\>\<n| \otimes
\left[ \begin{matrix} 0 & 0 \\ e^{\im \mu_n} r_n & e^{\im \tau_n} s_n \end{matrix} \right],
\label{eq6.1a}
\end{eqnarray}
where $r_n , s_n \ge 0$ and $\mu_n, \nu_n, \sigma_n , \tau_n \in  {\mathbb R}$,
and let $e^{ \im \delta_n}$  $(\delta_n \in {\mathbb R})$ be the determinant of 
\[
U_n = \left[ \begin{matrix} e^{\im \sigma_n} s_n & e^{\im \nu_n} r_n \\ 
 e^{\im \mu_n} r_n & e^{\im \tau_n} s_n\end{matrix} \right].
\]
$(U, \{\H_n \}_{n \in {\mathbb Z}})$ is a Szegedy walk if and only if the 
simultaneous equations 
\begin{equation}\label{eq6.1.4}
\theta_n - \theta_{n-1} -2  \lambda = \delta_n \quad  ({\rm mod} \  2\pi) 
\end{equation}
for all $n \in {\mathbb Z}$ and
\begin{equation}\label{eq6.1.5}
\theta_{n} - \lambda  = \mu_n  \quad ({\rm mod} \  \pi) \quad
{\rm when }\ r_n \neq 0 
\end{equation}
with respect to $ \lambda$ and $ \{ \theta_n \}_{n\in {\mathbb Z}}$
have a solution.
\end{theorem}

\proof
Since $e^{\im \delta_n}$ is the determinant of $U_n$,
\[
\delta_n = \sigma_n + \tau_n \quad ({\rm if} \ s_n \neq 0) \quad {\rm and} \quad
\delta_n = \mu_n + \nu_n +\pi\quad ({\rm if} \ r_n \neq 0)
\]
modulo $2\pi$.
Hence, equation \eqref{eq6.1.3} is calculated as
\[
\theta_n - \theta_{n-1} - 2\lambda = \sigma_n + \tau_n = \delta_n \quad ({\rm mod} \  2\pi).
\]
On the other hand, the first equation in \eqref{eq6.1.2} is equivalent to
\[
-2 \theta_n +2\lambda + 2 \mu_n =0 \quad ({\rm mod} \  2\pi),
\]
with the result that
\[
-\theta_n + \mu_n = \theta_n -2 \lambda - \mu_n \quad ({\rm mod} \  2\pi).
\]
Therefore, the second equation in \eqref{eq6.1.2} is calculated as
\[
\theta_n - \theta_{n-1} - 2\lambda = \mu_n + \nu_n + \pi = \delta_n \quad ({\rm mod} \  2\pi).
\]
Equation \eqref{eq6.1.5} is equivalent to the first equation in \eqref{eq6.1.2}.
Consequently, the simultaneous equations \eqref{eq6.1.2} and \eqref{eq6.1.3} have a solution
if and only if the simultaneous equations \eqref{eq6.1.4} and \eqref{eq6.1.5} have a solution.
\endproof

Another necessary and sufficient condition for a one-dimensional quantum walk
to be a Szegedy walk 
is easier to check in some cases.

\begin{corollary}\label{cor6.1.1}
Let $\{n_k \}_{k \in \Lambda} \subset {\mathbb Z}$ be numbers 
indexed by $\Lambda \subset {\mathbb Z}$ that satisfy $r_{n_k} \neq 0$ with $n_k < n_{k+1}$.
Suppose that $\Lambda \neq \emptyset$ and $0 \in \Lambda $.
A one-dimensional quantum walk $(U, \{\H_n \}_{n \in {\mathbb Z}})$ 
given by \eqref{eq6.1a}
is a Szegedy walk if and only if there exists $\eta \in {\mathbb R}$ such that
\[
\mu_{n_{k-1}} + \nu_{n_k } + \sum_{n=n_{k-1}+1}^{n_k -1} \delta_n 
=  \eta (n_k - n_{k-1} ) \quad ({\rm mod} \ \pi)
\]
for all $k \in \Lambda$ with $k-1 \in \Lambda$.
\end{corollary}

\proof
First, we assume that the simultaneous equations \eqref{eq6.1.4} and \eqref{eq6.1.5} 
have a solution $\{\lambda, \theta_n\}$.
By \eqref{eq6.1.4}
\begin{eqnarray}
\theta_{n_k} - \theta_{n_{k-1}} &=& 
\sum_{n=n_{k-1}}^{n_k-1} \theta_{n+1} - \theta_{n}
=2\lambda (n_k - n_{k-1} ) + \sum_{n=n_{k-1}}^{n_k-1} \delta_{n+1} \nonumber \\
&=& 2\lambda (n_k - n_{k-1} ) + \mu_{n_k}+ \nu_{n_k} +\sum_{n=n_{k-1}+1}^{n_k-1}   \delta_{n}
\quad ({\rm mod} \ \pi). \label{eq6.1.6}
\end{eqnarray}
Since $\theta_{n_k}$ and $\theta_{n_{k-1}}$ satisfy \eqref{eq6.1.5},
\[
\mu_{n_k} - \mu_{n_{k-1}} =  
2\lambda (n_k - n_{k-1} ) + \mu_{n_k}+ \nu_{n_k} +\sum_{n=n_{k-1}+1}^{n_k-1}   \delta_{n}
\quad ({\rm mod} \ \pi),
\]
with the result that
\[
\mu_{n_{k-1}} + \nu_{n_k } + \sum_{n=n_{k-1}+1}^{n_k -1} \delta_n 
=  -2\lambda (n_k - n_{k-1} ) \quad ({\rm mod} \ \pi).
\]

On the other hand, assume that there exists $\eta \in {\mathbb R}$ such that
\begin{equation}\label{eq6.1.7}
\mu_{n_{k-1}} + \nu_{n_k } + \sum_{n=n_{k-1}+1}^{n_k -1} \delta_n 
=  \eta (n_k - n_{k-1} ) \quad ({\rm mod} \ \pi)
\end{equation}
for all $k \in \Lambda$ with $k-1 \in \Lambda$.
Set $\lambda = -\eta/2$, $\theta_{n_0} = \mu_{n_0} + \lambda$, and
\[
\theta_n =
\left\{
\begin{array}{ll}
 \theta_{n-1} +2\lambda + \delta_n & ( n > n_0) \\
 \theta_{n+1} -2\lambda - \delta_{n+1} & ( n < n_0)
 \end{array}
\right.,
\]
inductively. Then, $\lambda$ and $\theta_n$ satisfy \eqref{eq6.1.4}.
Moreover, if $\theta_{n_{k-1}}$ with $n_{k-1}\ge n_0$ satisfies \eqref{eq6.1.5},
then $\theta_{n_k}$ also satisfies \eqref{eq6.1.5}. Indeed, by \eqref{eq6.1.6} and \eqref{eq6.1.7},
\begin{eqnarray*}
\theta_{n_k} &=& \theta_{n_{k-1}} +2\lambda (n_k - n_{k-1} ) +
 \mu_{n_{k}} + \nu_{n_k } + \sum_{n=n_{k-1}+1}^{n_k -1} \delta_n 
=
\theta_{n_{k-1}} +\mu_{n_k} - \mu_{n_{k-1}} \\
&=&  \lambda + \mu_{n_k}
\quad ({\rm mod} \ \pi).
\end{eqnarray*}
Similarly, 
 if $\theta_{n_{k}}$ with $n_{k} \le n_0$ satisfies \eqref{eq6.1.5},
then $\theta_{n_{k-1}}$ also satisfies \eqref{eq6.1.5}. Indeed, by \eqref{eq6.1.6} and \eqref{eq6.1.7},
\begin{eqnarray*}
\theta_{n_{k-1}} &=& \theta_{n_{k}} - 2\lambda (n_k - n_{k-1} ) 
- \mu_{n_{k}} - \nu_{n_{k} } - \sum_{n=n_{k-1}+1}^{n_k -1} \delta_n 
=
\theta_{n_{k}} - \mu_{n_k} + \mu_{n_{k-1}} \\
&=&  \lambda + \mu_{n_{k-1}}
\quad ({\rm mod} \ \pi).
\end{eqnarray*}
This completes the proof.
\endproof

As a special case of Corollary \ref{cor6.1.1}, we have the next corollary.

\begin{corollary}\label{cor6.2}
A one-dimensional quantum walk $(U, \{\H_n \}_{n \in {\mathbb Z}})$ given by \eqref{eq6.1a}
with $r_n \neq 0$ for all $n \in {\mathbb Z}$
is a Szegedy walk if and only if there exists $\eta \in {\mathbb R}$ such that
\[
\mu_{n-1} + \nu_n = \eta \quad ({\rm mod} \ \pi).
\]
\end{corollary}

When $r_n =0$ for all $n\in {\mathbb Z}$, a one-dimensional quantum walk is a Szegedy walk.

\begin{corollary}\label{cor6.3}
A one-dimensional quantum walk $(U, \{\H_n \}_{n \in {\mathbb Z}})$ given by \eqref{eq6.1a}
with $r_n = 0$ for all $n \in {\mathbb Z}$
is a Szegedy walk.
\end{corollary}
\proof
Set $\lambda= 0$, $\theta_0 =0$, and
\[
\theta_n = \left\{
\begin{array}{ll}
\theta_{n-1}  + \delta_n& (n \ge 1) \\
\theta_{n+1} - \delta_{n+1} &( n  \le -1)
\end{array}
\right. ,
\]
inductively. This is a solution of simultaneous equations \eqref{eq6.1.4}.
\endproof

Using these corollaries, we prove that every translation-invariant one-dimensional quantum walk
is a Szegedy walk.

\begin{corollary}
A translation-invariant one-dimensional quantum walk is a Szegedy walk.
\end{corollary}

\proof
If $r_n =0$ for all $n \in {\mathbb Z}$, then it is a Szegedy walk by Corollary \ref{cor6.3}.
If $r_n \neq 0$ for all $n \in {\mathbb Z}$, then $\mu_{n-1} + \nu_n$ is a constant,
because the quantum walk is translation invariant.
Therefore, it is a Szegedy walk by Corollary \ref{cor6.2}.
\endproof

Now, we consider some known models of one-dimensional quantum walks.

\begin{corollary}
A one-dimensional quantum walk $(U, \{\H_n \}_{n \in {\mathbb Z}})$ with $2$ coins
{\rm \cite{KRBD, EKO}},
i.e.,
\begin{eqnarray*}
U_{n-1, n} = 
\left[ \begin{matrix} a_+ & e^{\im \nu_+} r_+ \\ 0 & 0 \end{matrix} \right], \quad
U_{n+1, n} = 
\left[ \begin{matrix} 0 & 0 \\ e^{\im \mu_+} r_+ & d_+ \end{matrix} \right] \quad (n \ge 0) \\
U_{n-1, n} = 
\left[ \begin{matrix} a_- & e^{\im \nu_-} r_- \\ 0 & 0 \end{matrix} \right], \quad
U_{n+1, n} = 
\left[ \begin{matrix} 0 & 0 \\ e^{\im \mu_-} r_- & d_- \end{matrix} \right] \quad (n < 0) ,
\end{eqnarray*}
where $r_+, r_- >0$,
is a Szegedy walk if and only if 
\begin{equation}\label{eq6.1}
\mu_+ = \mu_-  \quad ({\rm mod} \ \pi), \qquad \nu_+ = \nu_- \quad ({\rm mod}\  \pi).
\end{equation}
\end{corollary}

\proof
By Corollary \ref{cor6.2},  $(U, \{\H_n \}_{n \in {\mathbb Z}})$ is a Szegedy walk if and only if
there exists $\eta \in {\mathbb R}$ such that
\[
\mu_+ + \nu_+ = \mu_- + \nu_+ = \mu_- + \nu_- = \eta \quad ({\rm mod} \  \pi).
\]
This condition is equivalent to
\[
\mu_+ = \mu_-  \quad ({\rm mod} \ \pi), \qquad \nu_+ = \nu_- \quad ({\rm mod}\  \pi).
\]
\endproof

Using Corollary \ref{cor6.2}, we have following two corollaries.

\begin{corollary}
The following quantum walk, considered in {\rm \cite{K2}}, 
\begin{eqnarray*}
U_{n-1, n} = 
\frac{1}{\sqrt{2}}
\left[ \begin{matrix}  e^{\im \omega_n} & 1 \\ 0 & 0 \end{matrix} \right], \quad
U_{n+1, n} = 
\frac{1}{\sqrt{2}}
\left[ \begin{matrix} 0 & 0 \\ 1 & -e^{-\im \omega_n} \end{matrix} \right] 
\end{eqnarray*}
is a Szegedy walk.
\end{corollary}

\begin{corollary}
The following quantum walk, considered in {\rm \cite{K3, KLS}},
\begin{eqnarray*}
U_{n-1, n} = 
\frac{1}{\sqrt{2}}
\left[ \begin{matrix}  1 & e^{\im \omega_n}  \\ 0 & 0 \end{matrix} \right], \quad
U_{n+1, n} = 
\frac{1}{\sqrt{2}}
\left[ \begin{matrix} 0 & 0 \\ -e^{-\im \omega_n} & 1 \end{matrix} \right] 
\end{eqnarray*}
is a Szegedy walk if and only if there exists $\eta \in {\mathbb R}$ such that
\[
- \omega_{n-1} + \omega_n = \eta \quad ({\rm mod}\  \pi).
\]
\end{corollary}

Using Theorem \ref{thm6.1}, we can prove that
a quantum walk of the Shikano-Katsura model  {\rm \cite{SK}} is a Szegedy walk.

\begin{corollary}
A quantum walk of the Shikano-Katsura model, i.e.,
\begin
{eqnarray*}
U_{n-1, n} = 
\frac{1}{\sqrt{2}}
\left[ \begin{matrix}  \cos (2\pi \alpha n) & -\sin (2 \pi \alpha n)  \\ 0 & 0 \end{matrix} \right], \quad
U_{n+1, n} = 
\frac{1}{\sqrt{2}}
\left[ \begin{matrix} 0 & 0 \\ \sin (2 \pi \alpha n) & \cos(2\pi \alpha n)  \end{matrix} \right] 
\end{eqnarray*}
is a Szegedy walk for any $\alpha \in {\mathbb R}$.
\end{corollary}

\proof
By the definition of $U$, $\mu_n$ and $\delta_n$ are $0$ or $\pi$ for all $n \in {\mathbb Z}$.
Set $\lambda =0$, $\theta_0=0$, and 
\[
\theta_n = \left\{
\begin{array}{ll}
\theta_{n-1} +\delta_n & (n \ge 1) \\
\theta_{n+1} -\delta_{n+1} & ( n \le -1)
\end{array}
\right. ,
\]
inductively.
This is a solution of 
\eqref{eq6.1.4} and \eqref{eq6.1.5}.
\endproof

\bigskip
\noindent
{\bf Acknowledgement.} 
The author is grateful to Etsuo Segawa for his helpful comments and suggestions of
the proofs of Theorem \ref{thm6.1} and Corollary \ref{cor6.1.1}.
The author also would like to express his gratitude to Akito Suzuki for his useful advice.
This work was supported by 
JSPS KAKENHI Grant Numbers 25800061.



\begin{thebibliography}{99}

\bibitem{ADZ}
Aharanov, L., Davidovidh, N., Zagury, N.:
Quantum random walks,
Phys. Rev. A {\bf 48}, 1687-1690 (1993)

\bibitem{A}
Ambainis, A.: Quantum walks and algorithmic applications,
Int. J. Quantum Inf. {\bf 1}, 507-518 (2003)

\bibitem{ABNVW}
Ambainis, A., Bach, E., Nayak, A., Vishwanath, A., Watrous, J.:
One-dimensional quantum walks,
Proc. 33th ACM Symposium of the Theory of Computing, 37-49 (2001)

\bibitem{EKO}
Endo, T., Konno, N., Obuse, H.:
Relation between two-phase quantum walks and the topological invariant,
arXiv:1511.04230.

\bibitem{GKD}
Goyal, S. K., Konrad, T., Di{\'o}si, L.:
Unitary equivalence of quantum walks,
Phys. lett. A {\bf 379}, 100-104 (2015)

\bibitem{G}
Gudder, S. P.:
Quantum Probability,
Academic Press, 1988.

\bibitem{HKSS1}
Higuchi, Yu., Konno, N., Sato, I., Segawa, E.:
Quantum graph walks I: mapping to quantum walks,
Yokohama Math. J. {\bf 59}, 33-55 (2013)

\bibitem{HKSS2}
Higuchi, Yu., Konno, N., Sato, I., Segawa, E.:
Quantum graph walks II: quantum walks on graph coverings,
Yokohama Math. J. {\bf 59}, 57-90 (2013)


\bibitem{HKSS3}
Higuchi, Yu., Konno, N., Sato, I., Segawa, E.:
Spectral and asymptotic properties of Grover walks on crystal lattices,
J. Funct. Anal. {\bf 267}, 4197-4235 (2014)

\bibitem{KRBD}
Kitagawa, T., Rudner, M. S., Berg, E., Demler, E.:
Exploring topological phases with quantum walks,
Phys. Rev. A {\bf 82}, 033429 (2010)


\bibitem{K1}
Konno, N.: Quantum random walks in one dimensional,
Quantum Inf. Process. {\bf 1}, 345-354 (2002)

\bibitem{K2}
Konno, N.: One-dimensional discrete-time quantum walks on random environment,
Quantum Inf. Process. {\bf 8}, 387-399 (2009)

\bibitem{K3}
Konno, N.:
Localization of an inhomogeneous discrete-time quantum walk on the line,
Quantum Inf. Process. {\bf 9}, 405-418 (2010)

\bibitem{KLS}
Konno, N., {\L}uczak, T., Segawa, E.:
Limit measures of inhomogeneous discrete-time quantum walks in one dimensional,
Quantum Inf. Process. {\bf 12}, 33-53 (2013)

\bibitem{SS}
Segawa, E., Suzuki, A.:
Spectral mapping theorem of an abstract quantum walk, 
arXiv:1506.06457.

\bibitem{SS2}
Segawa, E., Suzuki, A.:
Generator of an abstract quantum walk,
Quantum Stud. Math. Found.
Advance online publication. doi:10.1007/s40509-016-0070-1

\bibitem{SK}
Shikano, Y., Katsura, H.:
Localization and fractality in inhomogeneous quantum walks with self-duality,
Phys. Rev. E {\bf 82}, 031122 (2010)

\bibitem{Sz}
Szegedy, M.: Quantum speed-up of Markov chain based algorithms,
Proc. 45th Annu. IEEE Symp. Foundations Comput. Sci., 32-41 (2004)


\bibitem{V}
Venegas-Andraca, S. E.:
Quantum walks: a comprehensive review,
Quantum Inf. Process. {\bf 11}, 1015-1106 (2012)

\end{thebibliography}
\end{document}